\documentclass[sigconf,10pt,authorversion,nonacm]{acmart} %

\AtBeginDocument{%
  \providecommand\BibTeX{{%
    \normalfont B\kern-0.5em{\scshape i\kern-0.25em b}\kern-0.8em\TeX}}}

\usepackage{booktabs}
\usepackage{multirow}
\newcommand{\nomark}[0]{}

\newcommand{\tofab}{T_{A\leftrightarrow B}}
\newcommand{\tofat}{T_{A\leftrightarrow L}}
\newcommand{\tofbt}{T_{B\leftrightarrow L}}

\newcommand{\hattofab}{\hat{T}_{A\leftrightarrow B}}

\newcommand{\tofxy}{T_{X\leftrightarrow Y}}
\newcommand{\hattofxy}{\hat{T}_{X\leftrightarrow Y}}

\usepackage{xspace}
\newcommand{\n}[1]{\textcolor{blue}{#1}}

\renewcommand{\n}[1]{#1}

\newcommand{\tdoa}{TD}

\usepackage{subcaption}

\newcommand{\twr}{DS-TWR\xspace}
\newcommand{\td}{DS-TDoA\xspace}

\usepackage{tikz}

\definecolor{tab-red}{HTML}{d62728}
\definecolor{tab-blue}{HTML}{1f77b4}
\DeclareMathSymbol{\shortminus}{\mathbin}{AMSa}{"39}

\begin{document}

\title[Precise Ranging: Modeling Bias and Variance of Double-Sided Two-Way Ranging with TDoA Extraction under Multipath and NLOS Effects]{Precise Ranging: Modeling Bias and Variance of Double-Sided Two-Way Ranging with TDoA Extraction under Multipath and NLOS Effects}

\author{Patrick Rathje}
\email{pra@informatik.uni-kiel.de}

\affiliation{%
  \institution{Kiel University}
  \country{Germany}
}

\author{Olaf Landsiedel}
\email{ol@informatik.uni-kiel.de}

\affiliation{%
  \institution{Kiel University}
  \country{Germany}
}

\renewcommand{\shortauthors}{P. Rathje, O. Landsiedel}

\begin{abstract}
Location-based services such as autonomous vehicles, drones, and indoor positioning require precise and scalable distance estimates. The bias and variance of range estimators inherently influence the resulting localization quality. In this work, we revisit the well-established Double-Sided Two-Way-Ranging (\twr) protocol and the extraction of timing differences (\td) at devices overhearing \twr. Under non-line-of-sight (NLOS) and multipath effects, we analytically derive their bias and variance.
Our proposed model reveals that \twr retains half the variance than anticipated while \td comprises roughly a five-fold increase in variance. We conduct numerical simulations and experimental deployments using Ultra-Wideband (UWB) devices in a public testbed. Our results confirm the adequacy of our model, providing centimeter-accurate predictions based on the underlying timestamping noise with a median $R^2$ score of 77\% (30\% IQR). We find that both \twr and \td exhibit reduced variance when response times are symmetric. Our experimental results further show that double-sided variants exhibit less error and variance compared to Carrier Frequency Offset (CFO)-based single-sided methods.
\end{abstract}


\maketitle

\section{Introduction}

\paragraph{Context}
Whether for autonomous vehicles, the operation of drones, or real-time indoor localization, measuring distances with high precision and accuracy is an essential building block for location-dependent services~\cite{8692423}.
A prominent method is the Alternative-Double-Sided Two-Way-Ranging protocol (denoted as \twr), which estimates the distance based on the time of flight of wireless transmissions~\cite{neirynck2016alternative}. Its calculation allows mitigation of relative clock drifts based on a two-way message exchange, providing decimeter-accurate ranging on Ultra-Wideband (UWB) devices~\cite{neirynck2016alternative}.
Yet, the \twr protocol requires active participation, restricting its scalability in dense deployments due to limited channel capacity. Accordingly, recent works propose overhearing the \twr protocol on listening devices~\cite{horvath2017passive, rathje2022time, 10044978, puatru2023flextdoa, yang2022vuloc, moron2022towards, zhang2023self, chen2022pnploc}: overhearing devices infer the time difference of arrival (TDoA), i.e., the difference in distances to active devices. %
The retrieved TDoA information (denoted as \td from now on) can subsequently be used in localization algorithms, reducing the number of required \twr exchanges.

\begin{figure}
    \centering
    \includegraphics[width=\linewidth]{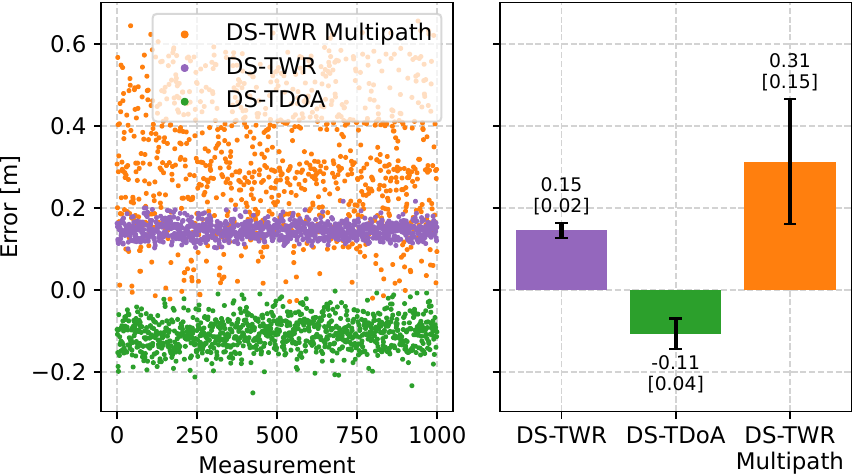}
    \caption{
    Exemplary measurements taken using Ultra-wideband devices show bias and variance in distance estimation. Bias and variance, often caused by multipath and NLOS propagation, directly translate to errors and uncertainties in positioning. Despite being ubiquitous ranging techniques, the state-of-the-art does not sufficiently model bias and variance in \twr and \td, especially in multipath and NLOS settings.
    \label{fig:raw}
    }
\end{figure}

\paragraph{Motivation}
Fundamentally, \twr measures the time of flight of radio transmissions, which travel at the speed of light. Consequently, the protocol demands precise timestamping capabilities of the underlying hardware.
Yet, the raw timestamps suffer from inherent noise caused by miscalibration, non-line-of-sight, or multipath conditions, impeding measurement precision.
Although \twr is widely adopted and deployed, its theoretical foundation is limited. Most works model noise on the level of the estimation rather than its relation to the underlying noisy timestamps. 
Furthermore, the lack of a solid theoretical foundation has led to speculative assumptions about the variance of \td, ranging from being comparable to DS-TWR~\cite{zhang2023self} to exhibiting twice as much variance~\cite{10044978}: This uncertainty underscores the need for a deeper understanding of timestamping noise's impact on measurement quality.

\paragraph{Challenges}
Even though the \twr protocol only consists of three messages between two parties, its inherent non-linearity caused by different clock frequencies complicates any theoretical analysis.
Moreover, the algorithm calculates durations based on noisy timestamps and employs them simultaneously for clock drift mitigation and range estimation, especially critical for complex NLOS or multipath scenarios. However, some durations are based on the same timestamps and share timestamping errors.
Hence, an accurate analysis requires meticulous tracking of those individual errors to capture this interdependency. The analysis of \td is even more intricate: it introduces another listening device with its inherent clock drift to the base \twr protocol and involves three additional (and noisy) reception timestamps recorded on that overhearing device.

\paragraph{Related Work}
Several works evaluate \twr's performance in practical scenarios~\cite{lian2019numerical, ledergerber2017ultra}; only a few investigate the variance of the \twr protocol theoretically. 
For one, Navr{\'a}til and Vejra{\v{z}}ka analyze the bias and variance of the ranging scheme using Taylor-Series approximation~\cite{navratil2019bias}. Their work employs Monte-Carlo simulations for verification but provides no practical evaluation.
Our work on the \twr protocol advances in \n{three} ways: For one, we exploit the shared nature of timestamps (and their errors) in our analysis. The result is a more accurate approximation, indicating that the theoretical variance is even lower than predicted in their work. \n{Further, our model extends to complex error distributions, commonly encountered in NLOS or multipath propagation scenarios, as displayed by Figure~\ref{fig:raw}}. 
Finally, in contrast to their work, our work verifies the analytical models using experimental results from testbed deployments.

For the \td operation, to our knowledge, our work is the first to assess its variance analytically.
Listed by Table~\ref{tab:related_work}, existing works commonly focus on the localization. While they diverge and disagree in their anticipated and measured variance for \td~\cite{zhang2023self,10044978}, we lay out the theoretical foundation and jointly verify the models for \twr and \td using numerical simulations and experimental results.%

\paragraph{Contributions}
This work presents a model for the established \twr protocol and its \td extension with regard to timestamping noise and response delays in \n{line-of-sight as well as complex non-line-of-sight and multipath environments.}
We analytically derive the expected bias and variance for \twr and \td operations. Our derivations track the underlying errors, which appear to counterbalance each other due to the shared nature of timestamps in the protocols.
\n{Based on our findings, despite the non-linear drift mitigation, any underlying noise bias caused, e.g., by NLOS and multipath effects, affects the protocols' means only linearly or may even counterbalance for \td.
Moreover, the influence of timestamping noise on the protocols' variance depends on the difference in delay times in the double-sided protocol: the least variance can be expected for symmetric response delays.
We verify our results using numerical simulations and a testbed deployment with Ultra-Wideband devices~\cite{molteni2022cloves}.
In addition, we compare the double-sided approach with relative drift mitigation based on clock frequency offset (CFO)~\cite{8555809}. Our experimental results show that bias and variance in \twr and \td remain low even under asymmetric delays. At the same time, CFO-based methods commonly result in a rise in variance and bias for longer protocol durations.}

The contributions of this work are as follows:
\begin{enumerate}
    \item By analytically deriving the bias and variance of both \twr and \td under NLOS and multipath effects, we provide the theoretical foundation for the well-established and ubiquitous ranging techniques.
    \item Under identical timestamping variance, \twr has half the variance than anticipated while \td exhibits a five-fold increase in variance compared to an active \twr exchange. The variances can be minimized using comparable response times.
    \item Using numerical simulations and experiments on UWB devices in a public testbed, we verify the adequacy of our analytical model with $R^2$ scores of up to 98\%. Our comparison to CFO-based variants indicates reduced bias and variance for double-sided variants. %
    \item We provide full access to our simulation code, firmware, and data, accompanied by its processing scripts.~\footnote{\url{https://github.com/ds-kiel/dstwr-variance}}
\end{enumerate}

\paragraph{Outline}
This work is organized as follows: After the preliminaries in Section~\ref{sec:background}, Section~\ref{sec:model} provides the measurement model \n{and mathematical notation; Section~\ref{sec:analysis} holds our analysis.} After Section~\ref{sec:numerical-results} validates our model in simulations, Section~\ref{sec:experimental-results} verifies it in a testbed deployment and compares bias and variance between double-sided and CFO-based approaches. Section \ref{sec:related-work} summarizes related work, and Section \ref{sec:conclusion} concludes this work.

\begin{table*}
    \centering
    \caption{Existing works analyze TWR protocols or  propose hybrid systems. This work focuses on the fundamental ranging performance of \twr and \td and includes NLOS/multipath propagation.\label{tab:related_work}}
    \begin{tabular}{lcccccccl}
        \toprule
        \multirow{2}{*}[-2pt]{Work} & \multicolumn{4}{c}{Model} & \multicolumn{3}{c}{Validation} & \multirow{2}{*}[-2pt]{Notes} \\
        \cmidrule(lr){2-5} \cmidrule(lr){6-8}
         & Bias & Var. & TDoA & NLOS & Sim. & Exp. & NLOS &\\
        \midrule
        Lian Sang et al.~\cite{lian2019numerical} & \checkmark & \nomark  & \nomark & \nomark & \checkmark & \checkmark & (\checkmark) & Comparison of TWR variants \\
        Shalaby et al.~\cite{shalaby2022reducing} & \checkmark & \checkmark & \nomark & \nomark & \checkmark & \checkmark & \nomark & Different DS-TWR protocol \\ 
        Navr{\'a}til \& Vejra{\v{z}}ka~\cite{navratil2019bias} & \checkmark & \checkmark & \nomark & \nomark & \checkmark & \nomark & \nomark & Theoretical DS-TWR analysis\\
        PnPLoc~\cite{chen2022pnploc} & \checkmark & \nomark &  \checkmark & \nomark & \nomark & (\checkmark) & \nomark & Focus on full system\\
                Zhang et al.~\cite{zhang2023self} & \checkmark & \nomark & \checkmark & \nomark & \nomark & (\checkmark) & \nomark &  Focus on full system \\
        Chiasson et al.~\cite{10044978} & \checkmark & \nomark & \checkmark & \nomark & \nomark & (\checkmark) & (\checkmark) & Only localization error \\
        VULoc~\cite{yang2022vuloc} & \checkmark & \nomark & \checkmark & \nomark & \nomark & (\checkmark) & (\checkmark) & Only localization error \\
        FlexTDoA~\cite{puatru2023flextdoa} & \checkmark & \nomark & \checkmark & \nomark & \nomark & \checkmark & (\checkmark) & CFO-Based with higher variance\\
        \textbf{This Work} & \checkmark & \checkmark & \checkmark & \checkmark & \checkmark & \checkmark &  \checkmark & Focus on underlying protocols \\ 
        \bottomrule
    \end{tabular}   
\end{table*}

\section{Background}\label{sec:background}
First, this section recapitulates the standard two-way ranging protocol and then summarizes the main idea of the Time Difference of Arrival extraction by overhearing a TWR exchange.

\subsection{Two-Way-Ranging}
By measuring the time-of-flight (ToF) of wireless transmissions, devices estimate inter-device distances, which are subsequently employed in localization algorithms. In line-of-sight (LOS) conditions, the primary source of error resides in the relative clock drifts of the two devices engaging in the measurement, as even slight deviations contest the resulting estimation accuracy: 
Every nanosecond of error in the estimate translates to approximately $30$~cm of error in distance.
Mitigation is provided by the Double-Sided Two-Way Ranging (TWR) protocol. This two-way message exchange allows two active devices to gauge the time of flight irrespective of clock offsets by comparing relative time intervals. Assuming a stable drift throughout the protocol execution, relative drifts can be approximated by comparing the overall execution time on both devices to reckon the relative drift~\cite{neirynck2016alternative}. %
Alternatively, as timestamps are generally provided by the radio clock, relative drifts can be estimated using the carrier frequency offset (CFO) of transmissions. Since the carrier frequency directly reflects the underlying clock rate, a receiver can measure the frequency offset to its own carrier frequency, estimating the relative drift~\cite{8555809}.
Due to their clock drift mitigation and support for asymmetric response delays, TWR protocols, both in their Double-Sided and CFO variants, are well-studied in experimental scenarios~\cite{lian2019numerical, ledergerber2017ultra}. %

\subsection{Time Difference of Arrival}
As the TWR protocols require active participation, the available channel capacity limits the number of active devices.
Consequently, scalable solutions record the time difference of arrival (TDoA) between receptions on overhearing devices. The resulting TDoA information enables overhearing devices to position themselves along hyperbolas with the active devices as focal points.
\n{Following this principle, works like SnapLoc~\cite{grossiwindhager2019snaploc}, Chorus~\cite{8732570} or TALLA~\cite{8911790} enable scalable TDoA-based tag UWB localization. TDoA approaches, however, require tight clock synchronization of the active devices.   
}

As the \twr protocol inherently relates timestamps to a common clock by mitigating clock drift and offsets (for the duration of the protocol), the combination of \twr and \td enables overhearing devices to estimate the TDoA without prior synchronization.
Such a combination is especially appealing in cooperative localization systems where mobile nodes act as anchors for other devices~\cite{buehrer2018collaborative}. %
\n{PnPLoc~\cite{chen2022pnploc}, for example, demonstrates how \twr and \td can be combined effectively using UWB to improve the accuracy and scalability of a system: Anchors with known positions execute the \twr protocol while passive tags rely on extracted TDoA information for positioning.}
However, several other works propose extracting TDoA information from a TWR exchange without a standard definition. The works differ in their derivations and use cases.
In some works, the mobile device performs TWR with active anchors while other anchors overhear the exchange and extract TDoA information~\cite{7993831, 8555809, horvath2017passive, s19245415, 9309999, 10044978}.
In others, like PnPLoc, the respective mobile device remains silent and instead extracts TDoA information from ongoing TWR between active anchors~\cite{chen2022pnploc, 10044978, yang2022vuloc, moron2022towards, zhang2023self}. Note that the underlying TDoA extraction is similar in both cases.

\section{A Model for \twr \& \td}
\label{sec:model}
This section introduces our underlying measurement assumptions and models an exemplary two-way ranging exchange with overhearing TDoA extraction. We first model an ideal exchange between devices and then add hardware-specific factors such as clock drift and reception timestamping noise to the model. The section then continues with the mitigation of relative clock drifts.
We end this section by discussing CFO-based mitigation and multipath effects.
The subsequent section then derives and analyzes the expected error and the variances for both \twr and \td.

\begin{figure}
    \centering
\includegraphics[width=0.9\linewidth, clip]{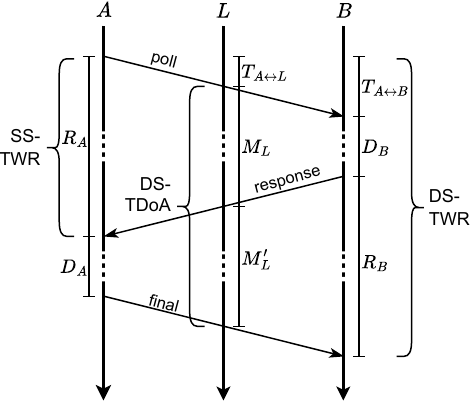}
    \caption{Message exchange in a Two-Way-Ranging Protocol with Overhearing: An overhearing node $L$ determines the difference in time-of-flight to $A$ and $B$ while these conduct active ranging using TWR.
    }
    \label{fig:twr}
\end{figure}

For the model definition, let $A, B$ denote two active nodes. Without loss of generality, we let $A$ initiate the active ranging process with $B$.
Further, let $L$ denote one of the potentially many devices that overhear this ranging process (e.g., some anchor or tag), recording the local reception timestamps of packets from $A$ and $B$.

We assume node movement is negligible during this operation, which generally executes within $10ms$. Thus, even speeds of 100 km/h lead to movements in the order of centimeters.

\subsection{Two-Way-Ranging}
The traditional or Single-Sided Two-Way Ranging (SS-TWR) computes the distance based on one poll and one response message in each direction, measuring the overall round time for the exchange as well as the delay of the other party, see SS-TWR in Figure~\ref{fig:twr}. 
Double-Sided Two-Way Ranging (\twr) builds on two interleaved Single-Sided rounds: Each side initializes one round with a message while the other responds.
As one message is shared between both rounds, three messages are exchanged, see DS-TWR in Figure~\ref{fig:twr}.
For the two active participants $A$ and $B$, we denote the actual round times (i.e., without any timestamping noise or clock drift) as $R_A$ and $R_B$ and, respectively, the response delays as $D_B$ and $D_A$. 
Let $\tofab$ denote the ground-truth time of flight (ToF) between participants $A$ and $B$. Without any errors or noise, for each single-sided poll-response exchange it holds that:
\begin{equation}
\tofab = \frac{R_A - D_B}{2}
\label{ss_tofab}
\end{equation}
Overall, we assume that messages contain all previous reception timestamps and their transmission timestamp, possibly using, for example, scheduled transmissions.

\subsection{Overhearing Time Difference of Arrival}
After introducing TWR, we now introduce how an overhearing party may extract information about the time difference of arrival (TDoA) and the difference of the arrival times as if $A$ and $B$ sent their messages simultaneously.
The TDoA is the difference in ToF from $L$ to $A$ and $L$ to $B$, respectively. We denote the ground-truth TDoA at node $L$ as $\tdoa$, for which holds:
\begin{equation}
\tdoa := \tofat-\tofbt
\label{tdoa}
\end{equation}
Assuming that $L$ receives the respective messages, it determines its local time difference between the poll issued by $A$ and the response message sent by $B$, which we denote as $M_L$ (cf. Figure~\ref{fig:twr}).
As the poll message from $A$ travels simultaneously to $L$ and $B$, the interval $M_L$ begins with a delay of $\tofat$ after the first message. It ends with a delay of $\tofbt$ after the response message from $B$. Hence, if we include the delay $D_B$, for $M_L$, the following holds:

\begin{equation}
M_L = (\tofab-\tofat) + D_B + \tofbt
\label{ss_ma_db_relation}
\end{equation}

The overhearing node also measures the duration $M_L'$ for the inverse exchange. Rearranging Equation \eqref{ss_ma_db_relation} and \eqref{tdoa} yields the basic TDoA to TWR relation:

\begin{equation}
\tdoa = \tofab + D_B - M_L 
\label{td_by_db}
\end{equation}

Based on Equation~\eqref{ss_tofab}, we convert the TDoA equation to the 
measurable response delays and round durations:
\begin{equation}
\tdoa =  0.5 R_A + 0.5 D_B - M_L
\label{td_by_meas}
\end{equation}

\subsection{Measurement Model}
After defining the ground-truth measurements, we now extend our model to include inherent noise and error sources that affect the quality of two-way ranging.
This work focuses on the noise stemming from reception timestamping covering both line-of-sight and non-line-of-sight (NLOS) as well as complex multipath scenarios.
In addition, we analyze the effect of multipath propagation in our numerical and practical evaluations in sections \ref{sec:numerical-results} and \ref{sec:experimental-results}.

\subsubsection{Timestamping Noise}
Reception timestamps are inherently noisy, and their quality degrades due to factors such as a low signal-to-noise ratio or multipath effects. \n{Typical causes are:
(1) Uncertainty in the timestamp due to fundamental noise in the signal recovery process, such as Gaussian noise of up to $150$~ps in the popular DW1000 UWB module~\cite{6881189}.
(2) Hardware-dependent influences like antenna delays~\cite{10160769, rathje2023aladin} and biases caused by the strength of the received signal ~\cite{note2019antenna,ledergerber2017ultra}.
And, (3) biases from multipath reflections or non-line-of-sight (NLOS) propagation~\cite{gallacherinsight, stocker2023applying}.}

\n{
Hence, the recorded reception timestamps contain inherent noise, affecting the quality of measurements and, hence, the quality of any derived positional estimate (cf. Figure~\ref{fig:raw}).

To accommodate this inherent noise, we make a key assumption: the timestamps of received packets are shifted by iid, additive noise. 
This timestamping noise affects all devices' receptions, i.e., the active participants $A$ and $B$ and the passive listener $L$. As $L$ determines the duration of $M_L$ (and $M_L'$) using two receptions, $M_L$ builds on two noisy receptions (one shared with $M_L'$). 
Moreover, the reception timestamping noise is generally not symmetric and increases with distance and a low signal-to-noise ratio~\cite{buehrer2018collaborative}. We hence denote the standard deviation of reception noise on, e.g.,  $B$'s receptions from $A$'s packets as $\sigma_{AB}$ and denote its mean as $\mu_{AB}$. We assume timestamping errors with finite variance distribution as follows:}
\n{\begin{align}
\varepsilon_{poll}, \varepsilon_{final} &\overset{\mathrm{iid}}{\sim} (\mu_{AB}, \sigma_{AB}^2)\\
\varepsilon_{resp} &\overset{\mathrm{iid}}{\sim} (\mu_{BA}, \sigma_{BA}^2)
\end{align}}
For the overhearing device $L$, our model assumes the following reception errors:
\n{\begin{align}
\varepsilon_{L_1}, \varepsilon_{L_3} &\overset{\mathrm{iid}}{\sim} (\mu_{AL}, \sigma_{AL}^2)\\
\varepsilon_{L_2} &\overset{\mathrm{iid}}{\sim} (\mu_{BL}, \sigma_{BL}^2)
\end{align}}
\n{Note that our model does not assume a specific distribution. The underlying noise could be a simple Gaussian or a complex, multi-modal distribution caused by multipath propagation. Hence, this error definition broadly captures a variety of noise sources, including hardware miscalibration and larger biases caused by NLOS propagation. Nevertheless, we assume the errors are diminutive compared to the response delays.}

\subsubsection{Clock Drift}

As clocks drift due to imperfections, they skew timestamps during the ranging protocol. Hence, the recorded durations deviate from the actual ones, requiring proper correction.

We denote the clock drift factor for a node $A$ as $k_A$. Note that the ranging protocol builds on intervals so that we can neglect any absolute clock offsets. As those drift factors mainly depend on the hardware and external factors, such as temperature, which are unlikely to change during the short execution time in the order of tens of milliseconds, we assume constant drift factors in our model.

We denote $X^Y$ as the skewed value of $X$ according to the clock of $Y$, i.e.: 
$$X^Y := k_Y X$$
We further assume that absolute deviations in a clock's frequency are negligible, i.e., a single clock measuring the time of flight introduces a minimal error.
For example, UWB devices, as standardized by IEEE 802.15.4~\cite{9144691}, are limited to clock drifts of less than $20$~ppm, i.e., sub-mm errors for a distance of $100$~m.

However, the drift in response delays is prominent and requires proper mitigation as differences skew the measured time of flight accordingly: If the delay drifts considerably, it may even exceed the round duration, resulting in a negative distance estimate. %
Consequently, we include clock drifts in our model and provide proper correction in Section~\ref{sec:model-rel-clock-drift-mitigation}.

\subsubsection{Combined Model}
We define $\hat{X}^Y$ as the measured duration or computed value of $X$ as measured in $Y$'s clock. %
Devices record these measurements; hence, the measurements inherit the timestamping noise from the individual receptions and are further skewed by their respective clock.
Accordingly, $\hat{R}_A^A$ describes the duration of $R_A$ as measured on node $A$. Combining the delays and reception noises, we derive the following relations for the recorded durations:

\begin{equation}
\hat{R}_A^A = k_A (R_A + \varepsilon_{resp}) = R_A^A + \varepsilon_{resp}^A
\label{drifted_ra}
\end{equation}
\begin{equation}
\hat{D}_B^B= k_B (D_B - \varepsilon_{poll}) = D_B^B  - \varepsilon_{poll}^B
\label{drifted_db}
\end{equation}
\begin{equation}
\hat{M}_L^L= k_L (M_L - \varepsilon_{L_1} + \varepsilon_{L_2}) = M_L^L - \varepsilon_{L_1}^L + \varepsilon_{L_2}^L
\label{drifted_ml}
\end{equation}

Due to the double-sided exchange, our model also contains the respective measurements for the second exchange (cf. Figure~\ref{fig:twr}). 

\begin{equation}
\hat{R}_B^B = R_B^B + \varepsilon_{final}^B
\label{drifted_rb}
\end{equation}
\begin{equation}
\hat{D}_A^A = D_A^A - \varepsilon_{resp}^A
\label{drifted_da}
\end{equation}
\begin{equation}
\hat{M}_L'^L= M_L'^L - \varepsilon_{L_2}^L + \varepsilon_{L_3}^L
\label{drifted_ml2}
\end{equation}

It is important to note that reception noise is shared between certain measurements. For example, this holds for $\hat{R}_A^A$ and $\hat{D}_A^A$ which both share $\varepsilon_{resp}^A$. This concept of shared timestamps allows us to derive the bias and variance analytically and results in a more accurate model as compared to the work of Navr{\'a}til and Vejra{\v{z}}ka~\cite{navratil2019bias}, as we also show in our evaluation (cf. sections \ref{sec:numerical-results} and \ref{sec:experimental-results}).

\subsection{Relative Clock Drift Mitigation}\label{sec:model-rel-clock-drift-mitigation}
Our model includes significant inaccuracies stemming from the underlying hardware (or multipath effects). However, different device clock frequencies lead to an inherent relative drift, which contests the protocols' accuracy as the durations are magnitudes larger than the time of flight.
Consequently, to mitigate the effect of relative clock drift, the measured durations must be converted to a common clock, e.g., all durations are converted to $A$'s timeframe.

We denote relative drift factors as $\frac{k_A}{k_B}$. For the double-sided protocol, we assume symmetry between the \twr rounds (before noise), i.e., assuming $R_A + D_A = D_B + R_B$, (see Figure~\ref{fig:twr})~\cite{neirynck2016alternative}. Note that this assumption neglects any relative movement during the protocol's execution between the devices, matching our initial assumptions. The drift between the active participants is approximated as follows:
\begin{equation}
\frac{k_A}{k_B} \approx \frac{\hat{R}_A^A+\hat{D}_A^A}{\hat{R}_B^B+\hat{D}_B^B}
\label{drift_factor}
\end{equation}
Simultaneously, $L$ determines relative drift rates to both $A$ and $B$ as follows:
\begin{equation}
\frac{k_L}{k_A} \approx \frac{\hat{M}_L^L+\hat{M'}_L^L}{\hat{R}_A^A+\hat{D}_A^A} \quad\quad \frac{k_L}{k_B} \approx \frac{\hat{M}_L^L+\hat{M'}_L^L}{\hat{R}_B^B+\hat{D}_B^B}
\label{drift_factor_passive}
\end{equation}
These formulas only approximate the relative drift factor owing to the inherent noise in the reception timestamps. However, this approximation benefits the estimation, lowering the protocols' variance.
In addition, its approximation requires knowledge of $B$'s round time $\hat{R}_B^B$, which needs an additional transmission by $B$.
Alternatively, relative drift factors can be estimated based on the Carrier Frequency Offset (CFO) on message reception~\cite{8555809}. This approach requires one message less but affects bias and variance as examined in Section~\ref{sec:exp-cfo}.

With knowledge of the relative drift, active party $A$ corrects for the clock drift in its time of flight measurements as follows: 
\begin{align}
\hattofab^A &:= 0.5 \hat{R}_A^A - 0.5 \hat{D}_B^A 
= 0.5 \hat{R}_A^A - 0.5 \frac{k_A}{k_B} \hat{D}_B^B \label{tof_corrected} %
\end{align}

Note that plugging in the relative drift estimate from Eq.~\eqref{drift_factor} into Eq.~\eqref{tof_corrected} is equivalent to the common Alt-DS-TWR formula~\cite{neirynck2016alternative}.
In the same manner, node $L$ converts the received duration $\hat{R}_A^A$, $\hat{D}_B^B$ to $\hat{R}_A^L$ and $\hat{D}_B^L$, mitigating the effect of the relative clock drift in its local TDoA calculations:
\begin{align}
\hat{\tdoa}^L :=& 0.5 \hat{R}_A^L + 0.5 \hat{D}_B^L - \hat{M}_L^L \nonumber \\
=& 0.5 \frac{k_L}{k_A} \hat{R}_A^A + 0.5 \frac{k_L}{k_B} \hat{D}_B^B - \hat{M}_L^L
\label{tdoa_corrected}
\end{align}
This formula for \td directly estimates the TDoA from an \twr exchange.

While the formulas include mitigation for relative clock drifts, the effect of noise remains to be analyzed. This holds especially true since the correction factor for relative drift also contains timestamping noise. Section~\ref{sec:analysis} analyzes \twr and \td as defined by Eq.~\eqref{tof_corrected} and \eqref{tdoa_corrected}.

\subsection{CFO-Based Mitigation}\label{sec:cfo-extension}
Our model assumes that relative clock drifts are mitigated using the double-sided variant, but relative clock drifts can also be mitigated using the carrier frequency offset in the single-sided case~\cite{8555809}.
Based on Eq.~\eqref{tof_corrected} and Eq.~\eqref{eq:td_base_equation}, we hence define Single-Sided variants, i.e., SS-TWR~\cite{8555809} and SS-TDoA~\cite{puatru2023flextdoa}, which solely rely on the CFO estimation. 
While this approximation is noisy, it does not require the extra final message of the \twr protocol.
The \td protocol proposed in this work requires another data dissemination message containing the value of $\hat{R}^B_B$ to compute the timing difference on the listening node. Note that this message's timing does not affect the estimate, and it could reside in another \twr packet, i.e., as part of another ranging process. Alternatively, instead of relying on another message from $B$, $L$ can approximate the relative drift $\frac{k_L}{k_B}$ using the CFO estimation while using the Double-Sided correction for $\frac{k_L}{k_A}$. This combination results in a mixed method, which we denote as Mixed-TDoA.
We compare and evaluate these variants in Section~\ref{sec:exp-cfo}.

\subsection{Multipath\,and\,Non-Line-of-Sight Effects} 
\n{
Our model generalizes timestamping noise and does not assume a specific distribution for the underlying error. Hence, it remains valid even in complex multipath propagation scenarios with multimodal error distributions.
}

\n{
The following section analyses both protocols in terms of their mean and variance under this generalized noise model. For verification, Section~\ref{sec:numerical-results} models an NLOS/multipath propagation by combining LOS Gaussian distribution with a randomized NLOS bias, resulting in a bimodal error distribution.
Likewise, our experimental results in Section~\ref{sec:experimental-results} contain multipath effects.
}

\section{Bias \& Variance Analysis} \label{sec:analysis}
In our model, \twr and \td suffer from two sources of errors: inherent clock drift and noisy reception timestamps. While we assume the clock drifts to be constant during the short execution period, mitigating the relative clock drift remains crucial for precise estimation. 
\n{We now} analyze the effect of clock drift and the reception noises on the expected value and variance in the Double-Sided case, i.e., approximating the relative drift using Eq.~\eqref{drift_factor}, respectively Eq.~\eqref{drift_factor_passive} for TDoA. For \twr, this formulation is semantically equivalent to the traditional Alternative-DS-TWR formulation~\cite{neirynck2016alternative}. Our formulation, however, captures the non-linearity in the relative drift components and allows for our detailed analysis.

\subsection{\n{Analysis of \twr}}

 Assuming $R_A+D_A=R_B+D_B$, for the calculation, i.e., negligible movement during the execution, we can analyze the protocol as follows (for convenience, calculated in $B$'s clock):
\begin{align}
\hattofab^B =& 0.5 \frac{\hat{R}_B^B+\hat{D}_B^B}{\hat{R}_A^A+\hat{D}_A^A} \hat{R}_A^A - 0.5 \hat{D}_B^B \nonumber \\ 
=& 0.5 \frac{R_B^B+\varepsilon_{final}^B+D_B^B-\varepsilon_{poll}^B}{R_A^A+\varepsilon_{resp}^A+D_A^A-\varepsilon_{resp}^A} \hat{R}_A^A - 0.5 \hat{D}_B^B \nonumber \\ 
\approx& 0.5 R_A^B - 0.5 D_B^B \nonumber + 0.5\varepsilon_{resp}^B +0.5\varepsilon_{poll}^B \\ &+ 0.5(\varepsilon_{final}^B-\varepsilon_{poll}^B)\frac{R_A^A}{R_A^A+D_A^A}
\label{tof_derivation}
\end{align}
Appendix~\ref{appendix:tof_derivation} lists the complete derivation.
Note that this calculation discards $\varepsilon_{final}^B*\varepsilon_{resp}^A$ and $\varepsilon_{poll}^B *\varepsilon_{resp}^A$ since the squared noise is diminutive compared to $R_A^A+D_A^A$.

As $\varepsilon_{final}$ and $\varepsilon_{poll}$ have the same mean value, we can safely disregard their difference and the delay ratio for the expected value. We further neglect the absolute drift of a single clock for the usual range of, e.g., UWB devices of less than $100$~m, as $k_A, k_B, k_L \approx 1 \pm 20*10^{-6}$, i.e., sub-mm deviations. 
The expected value forms as follows:
\n{
\begin{align}
&\mathbb{E}[\hattofab^B] = 0.5 R_A^B - 0.5 D_B^B + \mathbb{E}[0.5\varepsilon_{resp}^B] + \mathbb{E}[0.5\varepsilon_{poll}^B] \nonumber \\
=& k_B(\tofab + 0.5 \mu_{BA} + 0.5 \mu_{BA}) \approx \tofab + 0.5 (\mu_{BA} + \mu_{AB})\nonumber 
\end{align}
}
\n{Despite using timestamps for clock drift mitigation, their underlying noise skews the expected value symmetrically by their respective noise means, i.e., $0.5 (\mu_{BA} + \mu_{AB})$. The expected estimate is approximately unbiased in the case of zero-mean timestamping noise. Remarkably, even in complex multipath propagation scenarios where signals travel different paths within a single exchange, their impact boils down to a simple bias of the expected mean.}

We can derive the variance, as $k_B^2 \approx 1$, as follows:
\begin{align}
&Var[\hattofab^B] = k_B^2*Var[\hattofab] \approx Var[\hattofab] \nonumber \\
&= 0.5^2 \sigma_{BA}^2 + 0.5^2 (\frac{R_A^A}{R_A^A+D_A^A})^2 \sigma_{AB}^2 
+ 0.5^2 (1 - \frac{R_A^A}{R_A^A+D_A^A})^2 \sigma_{AB}^2 \nonumber
\end{align}

\subsection{\n{Analysis of \td}}
Similarly, assuming symmetry for the overhearing device, i.e., $R_A + D_A = R_B + D_B = M_L + M_L'$, we can now derive the mean and variance for the passive listener. 
Plugging in Eq.~\eqref{drift_factor_passive} into Eq.~\eqref{tdoa_corrected} yields:
\begin{align}
& \hat{\tdoa}^L = 0.5 \frac{\hat{M}_L^L+\hat{M}_L'^L}{\hat{R}_A^A+\hat{D}_A^A} \hat{R}_A^A + 0.5 \frac{\hat{M}_L^L+\hat{M}_L'^L}{\hat{R}_B^B+\hat{D}_B^B} \hat{D}_B^B - \hat{M}_L^L \label{eq:td_base_equation}
\end{align}
Following the same procedure as in Eq.~\eqref{tof_derivation} for the first drift mitigated component, it holds that:
\begin{align}
 \frac{\hat{M}_L^L+\hat{M}_L'^L}{\hat{R}_A^A+\hat{D}_A^A} \hat{R}_A^A
=& \frac{M_L^L + M_L'^L + \varepsilon_{L_3}^L - \varepsilon_{L_1}^L}{R_A^A+D_A^A} \hat{R}_A^A \nonumber \\
=& \hat{R}_A^L + (\varepsilon_{L_3}^L - \varepsilon_{L_1}^L) \frac{\hat{R}_A^A}{R_A^A+D_A^A} \nonumber \\
\approx& R_A^L + \varepsilon_{resp}^L + (\varepsilon_{L_3}^L - \varepsilon_{L_1}^L) \frac{R_A^A}{R_A^A+D_A^A}
\label{eq:td_first_component}
\end{align}
We can then approximate the second component as follows 
(full derivation available in Appendix~\ref{appendix:tdoa_first_component}):
\begin{align}
\frac{\hat{M}_L^L+\hat{M}_L'^L}{\hat{R}_B^B+\hat{D}_B^B} \hat{D}_B^B 
\approx D_B^L-\varepsilon_{poll}^L  - \frac{\varepsilon_{final}^L-\varepsilon_{poll}^L}{R_B^B+D_B^B} D_B^B +  \frac{ \varepsilon_{L_3}^L- \varepsilon_{L_1}^L }{R_B^B+D_B^B} D_B^B \label{eq:td_second_component}
\end{align}
As the time of flight is usually minute compared to the round and delay times, we further streamline and assume:
\begin{equation}
    \frac{R_A^A}{R_A^A+D_A^A},\frac{D_B^B}{R_B^B+D_B^B} \approx \frac{D_B}{D_B+D_A}
\end{equation}
Then, plugging Eq.~\eqref{eq:td_first_component} and  Eq.~\eqref{eq:td_second_component}
into Eq.~\eqref{eq:td_base_equation} yields:
\begin{align}
\hat{\tdoa}^L =& 0.5 \frac{\hat{M}_L^L+\hat{M}_L'^L}{\hat{R}_A^A+\hat{D}_A^A} \hat{R}_A^A + 0.5 \frac{\hat{M}_L^L+\hat{M}_L'^L}{\hat{R}_B^B+\hat{D}_B^B} \hat{D}_B^B - \hat{M}_L^L \nonumber \\
\approx& 0.5 R_A^L + 0.5 D_B^L - M_L^L \nonumber\\
&+ 0.5 \varepsilon_{resp}^L -0.5\varepsilon_{poll}^L  - 0.5 (\varepsilon_{final}^L-\varepsilon_{poll}^L)\frac{D_B}{D_B+D_A} 
\nonumber \\
&+ \varepsilon_{L_1}^L - \varepsilon_{L_2}^L + (\varepsilon_{L_3}^L - \varepsilon_{L_1}^L) \frac{D_B}{D_B+D_A}
\label{eq:td_mitigated}
\end{align}
\n{Surprisingly, \td's expected mean does not simply inherit \twr's bias:}
\n{\begin{align}
\mathbb{E}[\hat{\tdoa}^L] =& k_L \mathbb{E}[\hat{\tdoa}] \approx \tdoa + 0.5 \mu_{BA} - 0.5 \mu_{AB} + \mu_{AP} - \mu_{BP} \nonumber
\end{align}}
\n{When $\mu_{BA}$ and $\mu_{AB}$ are of the same magnitude, %
they appear to negate each other. Hence, \td's bias might remain unaffected even if the active \twr path suffers from multipath effects. Regarding its variance, it increases as the estimation incorporates two additional receptions (with iid errors):}
\begin{align}
&Var[\hat{\tdoa}^L] = k_L^2 Var[\hat{\tdoa}] \approx Var[\hat{\tdoa}]
\nonumber \\
&\approx
0.5^2 \sigma_{BA}^2
            + 0.5^2 \sigma_{AB}^2 (\frac{D_B}{D_B+D_A}-1)^2
            + 0.5^2 \sigma_{AB}^2 (\frac{D_B}{D_B+D_A})^2
            \nonumber \\
            &\quad + \sigma_{BP}^2
            +\sigma_{AP}^2 (1-\frac{D_B}{D_B+D_A})^2
            + \sigma_{AP}^2 (\frac{D_B}{D_B+D_A})^2
\nonumber \\
&\approx Var[\hattofab] + \sigma_{BP}^2 
            + \sigma_{AP}^2 (1-\frac{D_B}{D_B+D_A})^2
            + \sigma_{AP}^2 (\frac{D_B}{D_B+D_A})^2 \nonumber
\label{tdoa_var}
\end{align}
\subsection{Variance Optimality}
For both the TWR estimation and the  TDoA extraction, we can expect the least variance in the case of $\frac{D_B}{D_B+D_A}\approx 0.5$, hence, $D_A \approx D_B$, thus symmetric response times (disregarding the time of flight). Using numerical simulations, Section~\ref{sec:numerical-results} analyzes this dependency in simulation while Section~\ref{sec:experimental-results} covers testbed experiments. \n{Assuming constant drift rates, delays do not affect the experienced bias. However, we find that \td experiences increased variance than \twr.}%

\section{Numerical Results}
\label{sec:numerical-results}
In this section, we verify our theoretical analysis using Monte-Carlo Simulations before discussing experimental results in Section~\ref{sec:experimental-results}.
We analyze the theoretical performance of our proposed \td scheme and compare it to the \twr approach, verifying our analytical model. We evaluate metrics such as the mean and the standard deviation of the estimated TWR distance and TDoA distance difference.
For this experiment, we assume identical reception noise for all LOS transmission paths, an assumption shared by the variance model of Navr{\'a}til and Vejra{\v{z}}ka~\cite{navratil2019bias}. Accordingly, we use their model as a baseline for comparison. Another analytical model by Shalaby et al. builds on a different TWR protocol and is, hence, not comparable~\cite{shalaby2022reducing}. \n{In addition to LOS paths, we introduce NLOS obstacles resulting in multipath scenarios.}

\subsection{Simulation Setup}
As described in Section~\ref{sec:model}, we simulate two active nodes that execute a TWR exchange and a third overhearing node that runs the TDoA extraction scheme. Devices mitigate the relative clock drift using Eq.~\eqref{drift_factor}.
We do not vary positions as the measurements do not depend on the actual position in this simulated setup. %
For each run, clock drifts are assumed to be constant, independent, and zero-centered, with a standard deviation of $10$~ppm.
\n{We define three obstacle positions%
; each impedes one of the three propagation paths (see Figure~\ref{fig:sim_scenario}).
Without an obstacle, a link experiences symmetric, zero-centered Gaussian reception noise as $\varepsilon_{LOS} \overset{\mathrm{iid}}{\sim} \mathcal{N}(0, \sigma_{RX}^2)$ with $\sigma_{RX} = 1ns$. However, if an obstacle is present, we simulate a bimodal noise distribution by introducing a $4ns$ NLOS/multipath bias with 50\% probability:%
$$\varepsilon_{NLOS} := \varepsilon_{LOS} + \beta 4ns \qquad \beta\overset{\mathrm{iid}}{\sim} Bern(0.5)$$ 
}%
\n{We further vary the ratio of response delays, i.e., $\frac{D_B}{D_B+D_A}$ from $0.001$ to $0.999$, as we expect an evident dependency according to our model.
Per ratio and LOS/NLOS combination, we simulate 2,000 independent exchanges. %
We compare the sample bias and variance to our analytical model predictions. For comparison, we provide the analytical variances for \twr in LOS scenarios as defined by Navr{\'a}til and Vejra{\v{z}}ka~\cite{navratil2019bias}.
Their model, however, does not include NLOS or multipath conditions and further does not include \td's bias or variance, for which this work is the first to propose an analytical model.}

\subsection{Simulation Results}

\begin{figure*}
\centering
\captionsetup[subfigure]{justification=centering}
\begin{subfigure}[b]{.2\linewidth}%
\centering\hspace*{6mm}\includegraphics[scale=0.9]{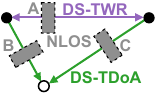}\vspace{3.6mm}\\
\includegraphics[scale=0.5]{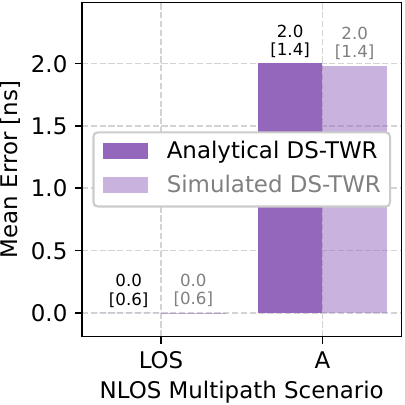}
    \caption{\twr under NLOS\label{fig:sim_scenario}\label{fig:sim_bias_twr}} 
  \end{subfigure}%
  \begin{subfigure}[b]{.4\linewidth}%
\centering\includegraphics[scale=0.5]{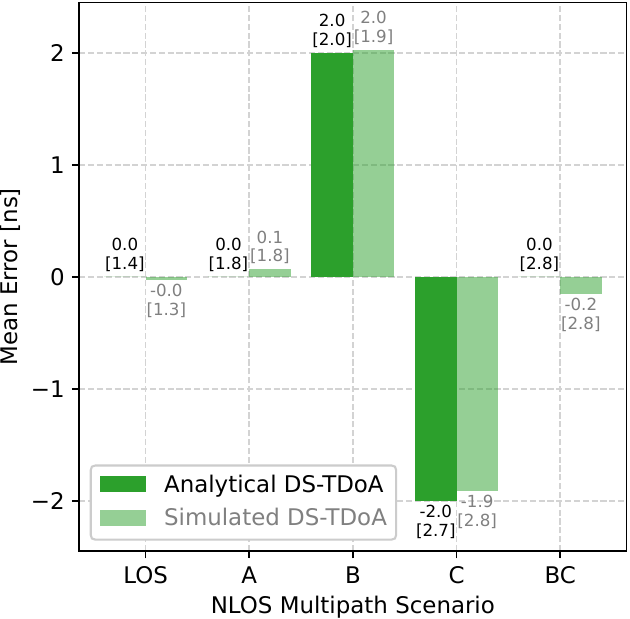}%
    \caption{\td  under NLOS\label{fig:sim_bias_td}}
  \end{subfigure}%
  \begin{subfigure}[b]{.39\linewidth}%
\centering\includegraphics[scale=0.5]{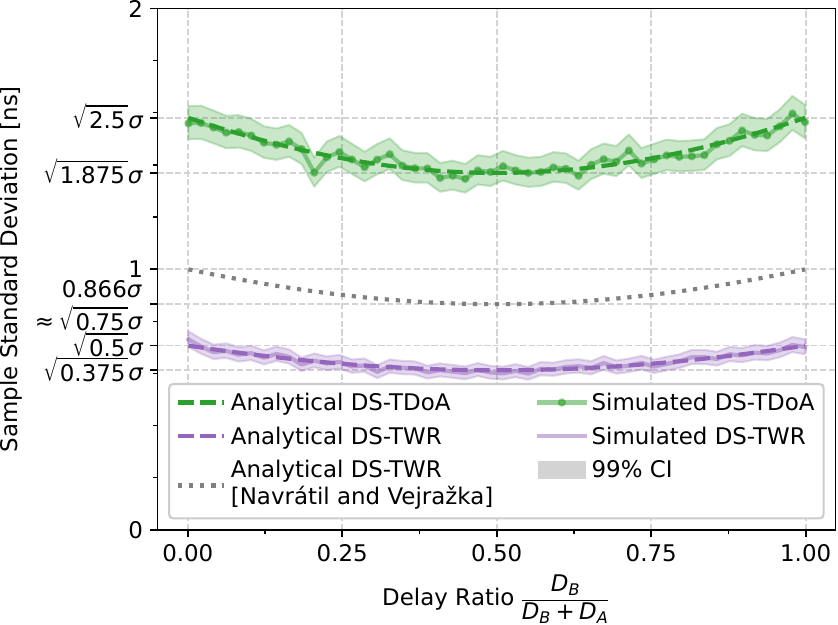}
    \caption{Delay Ratio vs. Variance in LOS\label{fig:sim_std_response_ratio}}
  \end{subfigure}%
\caption{\n{
Our analytical model predicts sample bias and standard deviation in Monte-Carlo simulations, even under NLOS effects.
It provides a more accurate estimate than existing variance models, which are limited to \twr in LOS conditions~\cite{navratil2019bias}.
We simulate NLOS effects for \twr (a) and \td (b). The standard deviation reduces when the delays of the active devices are symmetric. \td's scalability comes with an increase in variance (c). %
\label{fig:exp_simulation}}}
\end{figure*}

\n{
Figure~\ref{fig:sim_bias_twr} and Figure~\ref{fig:sim_bias_td} display the scenario and results for \twr and \td under  LOS and NLOS effects. For both protocols, our model closely predicts the bias and variance. Different delay ratios did not influence the bias; we only depict the symmetric delay scenario. Due to the assumed symmetry, an obstacle between the active devices only affects the variance, not the bias of \td.}

Figure~\ref{fig:sim_std_response_ratio} illustrates the effect of different delay ratios in an all-LOS scenario:
The experienced sample standard deviations match the theoretically derived ones.
At the same time, the comparison with the model by Navr{\'a}til and Vejra{\v{z}}ka~\cite{navratil2019bias} shows an increase in prediction accuracy by our model. Their work also employs Monte Carlo simulations, but each run samples noise from a covariance matrix, which does not preserve exact timestamps relations. Our simulations simulate the whole protocol exchange and capture the inter-dependency of noise in \twr, as some timestamps are shared in the standard protocol (see Section~\ref{sec:model}).
In addition, we do not record any noticeable effect on the sample standard deviation when we vary node drifts or the magnitude of the overall response delay.%

\section{Experimental Results}\label{sec:experimental-results}
This section verifies our theoretical and numerical results in a testbed deployment. For easy reproducibility, we execute our experiments on Cloves, a public testbed featuring Qorvo's DWM1001 UWB transceivers~\cite{molteni2022cloves}. After presenting our setup, we investigate and compare the \n{bias and} variance on actual hardware. Then, using the same experimental setup, we compare the performance of the double-sided variants with CFO-based approaches.

\subsection{Experimental Setup}
We deploy our firmware on 7 DWM1001 UWB transceivers in a (theoretically) line-of-sight scenario; see Figure~\ref{fig:exp_setup_testbed_layout}. \n{Yet, even this simple scenario contains multipath propagation.}

 Devices use channel 5 with a PRF of $64$~MHz and a preamble length of 128 symbols, a typical setup for short-range communication and ranging in UWB. During startup, devices load factory-calibrated antenna delays from OTP memory. 
On reception, devices mitigate the non-linear range bias based on the reported received signal strength using factory-provided correction values~\cite{nodes2014sources}. 
Devices log transmission and reception timestamps over a serial backbone with $40$~bit precision, i.e., with a resolution of $15.65$~ps, equivalent to $0.47$~cm. 
We skip the first 30 minutes of records to reduce the influence of startup drift~\cite{krvska2023stability}.

\n{In our experiments, nodes execute a series of back-and-forth exchanges with other nodes, one pair at a time, while the other nodes overhear their exchange.}
Each active pair exchanges up to $201$ messages with a constant delay of $750$us. We pick the subset of exchanged messages based on the desired overall protocol length and delay ratio.

Figure~\ref{fig:exp_setup_median_error} exemplary depicts the bias deviations of the pair 4-6 under varying delay ratios and durations of the protocol: \n{With increasing durations, estimates deviate from the initial estimates,} suggesting a change in relative drift rates, which naturally breaks our assumption of constant clock drifts.
As expected, for typical execution times of less than $40$~ms, the mean error remains steady and, simultaneously, unaffected by a change in the delay ratio due to the stable relative drift.

\subsection{Estimating Timestamping Noise}
Our proposed model builds on the intrinsic timestamping noise affecting \twr and \td protocols. The bias and variance of this noise are essential parameters. While we can control the noise in our numerical simulations, we must estimate those parameters in our deployment.

In our experiments, the devices reveal an error of up to $31$~cm compared to the ground truth distance. For our employed UWB devices, those biases commonly stem from missing calibration of innate delays in signal processing~\cite{10160769, rathje2023aladin} or the relative orientation of antennas, likely caused by non-isotropic antenna patterns~\cite{ledergerber2017ultra}. Further sources of error remain in multipath or NLOS effects, which NLOS classification can mitigate~\cite{stocker2023applying}.
\n{As the error sources are complex, we cannot precisely determine the underlying bias without any backbone for time synchronization. Instead, we compare \twr's 
  reported distance (sample mean) with available ground-truth distances. Assuming a symmetric bias, we reckon the underlying noise mean as follows:}
  $$\mu_{XY} = \mu_{YX} \approx \mathbb{E}[\hattofxy]  - \tofxy$$

\begin{figure*}
\centering
\captionsetup[subfigure]{justification=centering}
\begin{subfigure}[t]{.3\linewidth}
    \centering\includegraphics[width=0.9\linewidth]{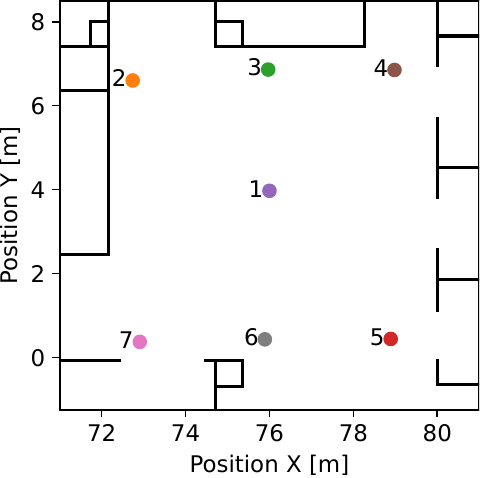}
    \caption{CLOVES installation with 7 Ultra-Wideband devices~\cite{molteni2022cloves} \label{fig:exp_setup_testbed_layout}} 
  \end{subfigure}
  \begin{subfigure}[t]{.34\linewidth}
    \centering\includegraphics[width=0.8\linewidth]{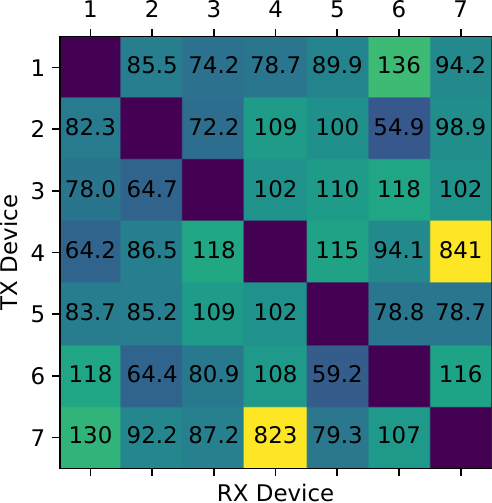}
    \caption{Estimated Timestamping Noise\\Standard Deviation [ps]\label{fig:exp_setup_rx_noise_sd}}
  \end{subfigure}
  \begin{subfigure}[t]{.3\linewidth}
    \centering\includegraphics[width=\linewidth]{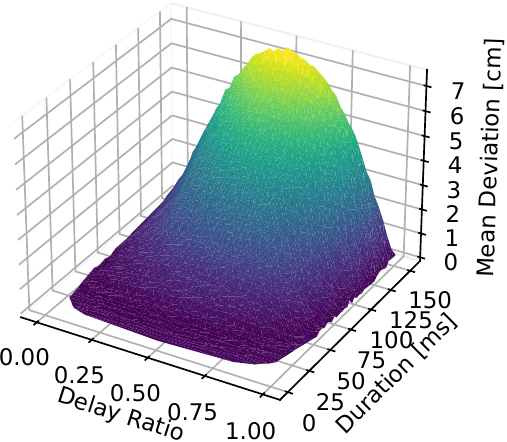}
    \caption{
    Influence of Delay Ratio and  Duration on \twr Bias (Pair 4-6)\label{fig:exp_setup_median_error}}
  \end{subfigure}
\caption{We run our experiments on Ultra-Wideband nodes in a public testbed (a).
Pair 4-7 suffers from multipath propagation, resulting in a high estimate of the underlying timestamping variance (b). Pairs reveal errors of up to $31$~cm compared to the ground truth distances using \twr. Longer protocol durations, in particular, potentially break our assumption of constant clock drifts and shift the distance estimate (c).\label{fig:exp_setup}
}
\end{figure*}

\n{For estimating the variance, without a backbone for clock synchronization, we rely on message exchanges for each transmission path:} We group reception timestamps into distinct $60$~ms windows (approximately $40$ receptions). Employing linear least squares, we estimate the relative drift and absolute offset between the transmitter's clock and the receiver's clock for each window. The resulting residual standard error gauges the standard deviation of the reception noise.
For each pair, we assume an underlying timestamping variance equal to the respective median of estimations. We empirically choose the $60$~ms duration, as only then does the estimate converge for all pairs.
Multiple transmission paths, i.e., due to multipath effects, are captured by the noise bias and variance in our model (cf. Figure~\ref{fig:raw}). Consequently, the multipath effects result in high timestamping variance. Figure~\ref{fig:exp_setup_rx_noise_sd} displays the respective timestamping deviation for all pairs, of which all (except for one multipath pair) fall into the typical range of around $100$ to $150$~ps in LOS~\cite{6881189}.

Note that we estimate timestamping noise under our assumption of constant drift. Hence, the estimated variance potentially contains various types of clock noise. Other approaches, such as Time Deviation, categorize different types of noise, providing additional insights into the underlying noise types~\cite{krvska2023stability}.

\subsection{Bias of \twr and \td}

\begin{table}
  \caption{\n{Predicting \td's Bias from \twr on Multipath-affected Node 7 (Excerpt)}\label{tab:bias_pre}}
  \begin{tabular}{crrr}
    \toprule
    Active Pair & Predicted\,[m] & Actual\,[m] & Error\,[m] \\
    \midrule

$4 \rightarrow 1$ & $0.445$ & $0.464$ & $-0.019$ \\$1 \rightarrow 4$ & $-0.445$ & $-0.461$ & $0.016$ \\
$4 \rightarrow 2$ & $0.340$ & $0.332$ & $0.008$ \\
$2 \rightarrow 4$ & $-0.340$ & $-0.348$ & $0.008$ \\
$4 \rightarrow 3$ & $0.436$ & $0.433$ & $0.003$ \\
$3 \rightarrow 4$ & $-0.436$ & $-0.448$ & $0.013$ \\
$4 \rightarrow 5$ & $0.301$ & $0.322$ & $-0.021$ \\
$5 \rightarrow 4$ & $-0.301$ & $-0.318$ & $0.016$ \\
  \bottomrule
\end{tabular}
\end{table}

\n{Our model predicts the resulting bias of both \twr and \td based on the underlying noise, as supported by our numerical results. We now set out to analyze the bias interplay between both protocols.}

\n{We estimate the timestamping bias per link from \twr using available ground-truth distances. We collect over $600$ measurements with symmetric response delays of $0.75$~ms. Averaging over all tested pairs, we obtain a mean absolute error of $8.2$~cm for raw \twr and $14.3$~cm for raw \td estimates. We predict \td's biases for all active pairs and listening devices based on the gauged noise.}

\n{Despite the larger errors for \td, our prediction reveals an average absolute error of only $1.3$~cm with a maximum deviation of $3$~cm. Table~\ref{tab:bias_pre} lists an excerpt of \td bias predictions.}

\subsection{Variance of \twr and \td}
\begin{figure*}
\centering
\captionsetup[subfigure]{justification=centering}
\begin{subfigure}[t]{.32\linewidth}
    \centering\includegraphics[width=\linewidth]{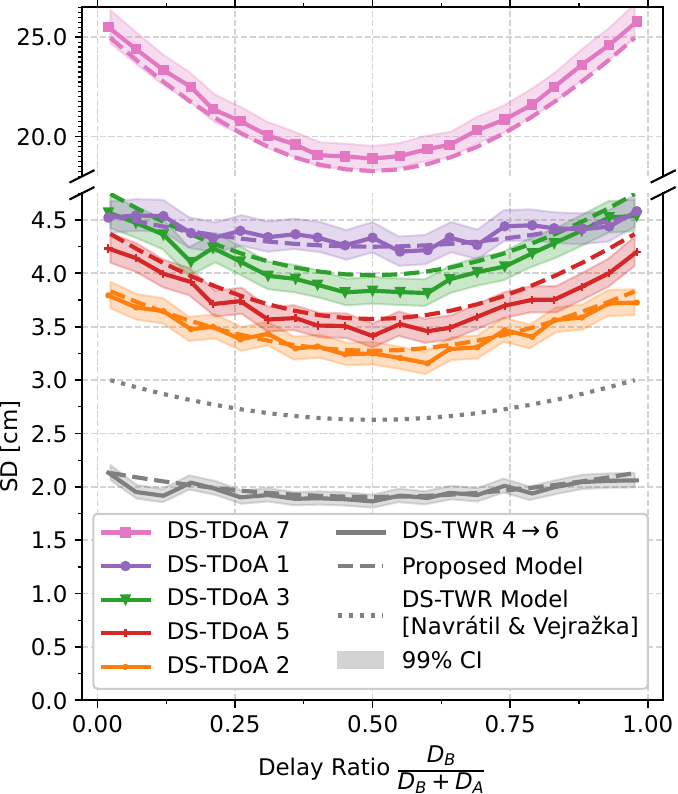}
    \caption{\twr pair 4-6 \label{fig:std_first}}
  \end{subfigure}
  \begin{subfigure}[t]{.32\linewidth}
    \centering\includegraphics[width=\linewidth]{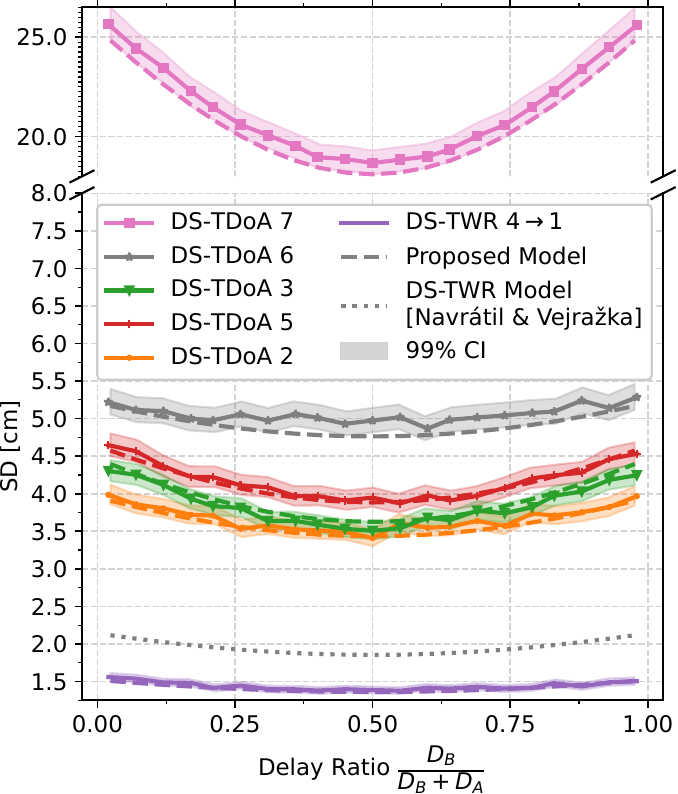}
    \caption{\twr pair 4-1, overhearing node 6 exhibits a low $R^2$ score of -1.8\label{fig:std_second}}
  \end{subfigure}
  \begin{subfigure}[t]{.32\linewidth}
    \centering\includegraphics[width=\linewidth]{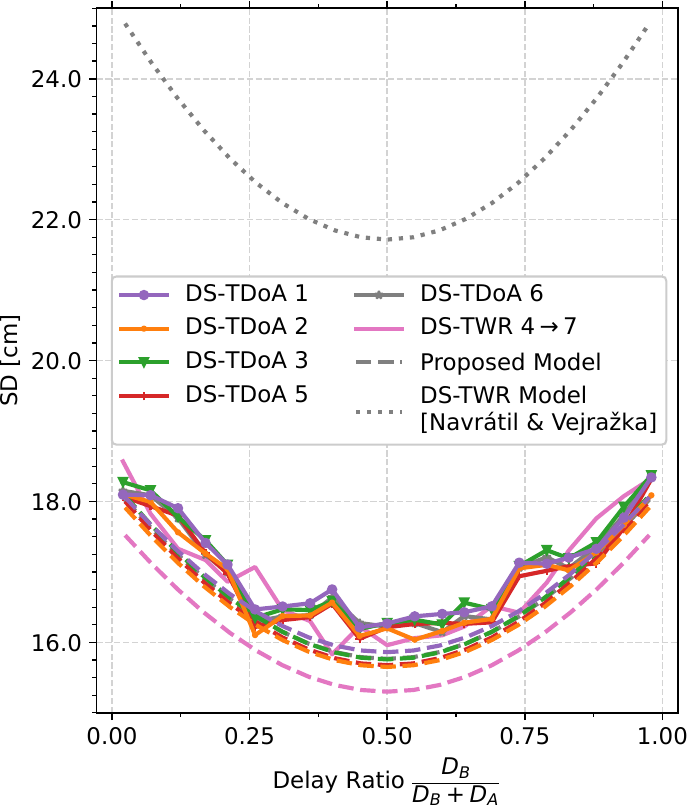}
    \caption{DS-TWR Pair 4-7 suffering from  multipath propagation\label{fig:std_third}}
  \end{subfigure}
\caption{
    Experimental results using UWB transceivers: A distinct pair of nodes executes the \twr protocol, and other devices overhear the exchange for \td information. Based on our estimated timestamping noise, our model predicts the resulting standard deviation for \twr and \td. The empirical data matches our analytical model for both protocols, even in the case of multipath effects. Our model provides closer predictions than existing \twr models and further captures the difference between \twr and \td variance.
    }
    \label{fig:std-fit}
\end{figure*}

We now compare our model's variance predictions with our testbed results, verifying its adequacy in a realistic scenario. 

Figure~\ref{fig:std-fit} illustrates an excerpt of our recorded sample variance of different \twr exchanges and \td extraction on all overhearing devices. We collect over 3,000 samples per ratio throughout ten independent runs and plot the overall sample standard deviation with their respective confidence intervals. We choose an overall protocol duration of $31.5$~ms, which is longer than usual protocol executions but allows us to explore extremely asymmetrical scenarios while maintaining the same variance level for double-sided protocols.

We employ our analytical model to predict the expected variance based on the estimated underlying timestamping variance. \n{As we estimate the underlying variance from the raw records, we predict variances for both protocols, visualized by dotted lines in Figure~\ref{fig:std-fit}.} %
We also illustrate the predictions for \twr according to Navr{\'a}til and Vejra{\v{z}}ka~\cite{navratil2019bias}.

Overall, our experimental findings indicate a trend towards our model's predictions. 
$R^2$ scores of up to $76.5\%$ for \twr and $98.2\%$ for \td support this finding, i.e., our model can explain much of the inherent variance with a median of 77\% (30\% IQR). %
Table~\ref{tab:r2_pred} summarizes all observed $R^2$ values for the active exchanges and their overhearing devices.

We observe lower $R^2$ scores in the case of multipath effects for the pair 4-7 (displayed in Figure~\ref{fig:std_third}). In addition, node 6 exhibits more noise than predicted when overhearing \twr between nodes 4 and 1 (cf. Figure~\ref{fig:std_second}). We suspect this pair is already drifting apart in the chosen timeframe, causing additional noise, as shown by Figure~\ref{fig:exp_setup_median_error}.

Nevertheless, in all cases, our predictions remain close to the empirical data for \twr, while the baseline model by Navr{\'a}til and Vejra{\v{z}}ka predicts more variance than what we observe. In contrast to existing models, our model can predict \td's variance, capturing the interplay between \twr and \td.

\begin{table}
  \caption{$R^2$ Prediction Scores in Testbed Deployment \label{tab:r2_pred}}
  \begin{tabular}{crr}
    \toprule
    Active Pair & DS-TWR $R^2$ & DS-TDoA $R^2$ \\
    \midrule
    $4\rightarrow 1$ & $0.60$ &             $0.77, 0.89, 0.95, \mathbf{\shortminus 1.5}, 0.91$\\
    $4\rightarrow 2$ & $0.55$ &             $0.73, 0.92, 0.89, 0.87, 0.90$\\
    $4\rightarrow 3$ & $0.17$ &             $0.78, 0.97, 0.60, 0.87, \mathbf{0.98}$\\
    $4\rightarrow 5$ & $0.77$ &             $0.93, 0.74, 0.78, 0.93, 0.95$\\
    $4\rightarrow 6$ & $0.45$ &             $0.65, 0.90, 0.69, 0.68, 0.91$\\
    $4\rightarrow 7$ & $\mathbf{-0.06}$ & $0.60,  0.56, 0.56, 0.62, 0.66$\\
  \bottomrule
\end{tabular}
\end{table}

\subsection{Comparison with CFO-Based Methods} \label{sec:exp-cfo}
The CFO-based approach estimates the relative clock drift by measuring the carrier frequency offset between the sending and the receiving device, and hence, it does not require a third message.
For short durations, the CFO-based approaches perform similarly to the double-sided variants regarding their mean error~\cite{8555809} while  FlexTDoA~\cite{puatru2023flextdoa} indicates increased variance for longer delays. In this experiment, we compare the variance of \twr and \td with the single-sided, CFO-based variants, i.e., SS-TWR and SS-TDoA. 

We let nodes 4 and 6 complete the double-sided protocol in $7.5$~ms with node 1 overhearing this exchange. We vary the response delay $D_B$, which defines the single-sided protocol's total duration and the double-sided variant's delay ratio. The double-sided protocol performs similarly to the longer execution times (cf. Figure~\ref{fig:std_first}).
We use the carrier integrator estimates recorded by the DW1000 UWB chip to mitigate relative clock drifts.
For the TDoA protocols, we introduce the Mixed-TDoA variant (see Section~\ref{sec:cfo-extension}), which offers a hybrid approach by combining CFO estimation for the responder's relative drift and the traditional double-sided mitigation for the initiator, overall saving one transmission. We do not employ any filtering to track the relative clock drift; all exchanges are estimated independently.

\begin{figure}
    \centering
    \includegraphics[width=1.0\linewidth, clip, trim=0.0cm 0cm 0.0cm 0cm]{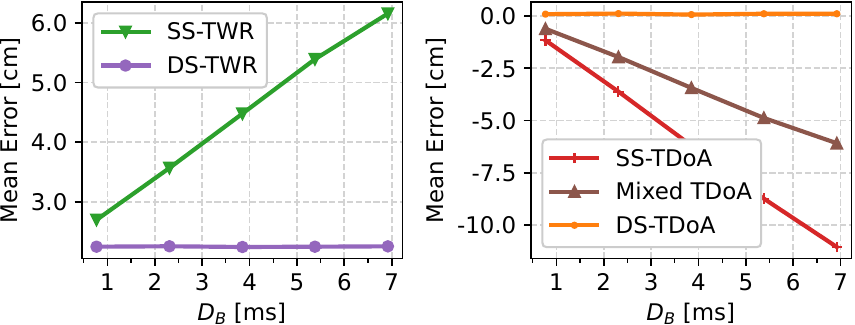}    \includegraphics[width=1.0\linewidth, clip, trim=0.0cm 0cm 0.0cm 0cm]{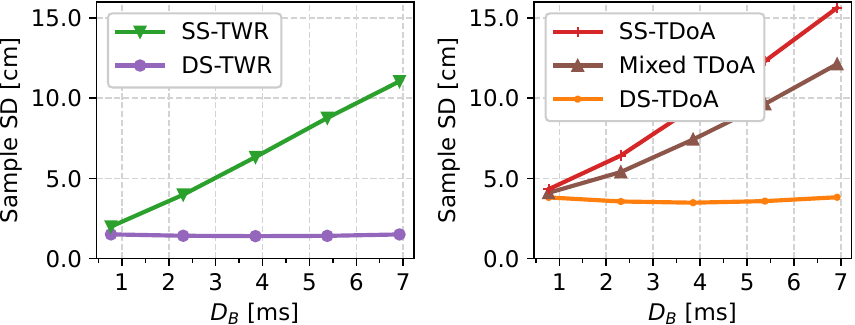}
    \caption{
        Comparing the bias and 
 variance of CFO-based and double-sided 
 protocols for different response delays:
        While the CFO-based methods require one message less, they exert increased bias and variance for longer delays. For overhearing devices, only the proposed double-sided approach maintains a low variance. The suggested Mixed-TDoA exerts higher variance than the double-sided method but requires one message less than \td. Yet, it shows less variance than the fully CFO-based alternative.\label{fig:cfo_comparison}
    } %
\end{figure}

Figure~\ref{fig:cfo_comparison} depicts the derived bias and standard deviation from our experiments. For small response delays, all approaches perform comparably in terms of their mean.
While the CFO-based variants generally require one message less, they also demonstrate an increase in variance with longer response delays.
Yet, for the shortest delay $D_B$ of $0.77$~ms, the single-sided variants only add around $0.5$~cm of additional noise on top of \twr and \td.
This result is consistent with analyses of CFO's quality on the same DWM1001 UWB device~\cite{krvska2023stability}: CFO-based estimation is recommended only for short response delays or when accuracy is not critical. \n{Note that the performance depends on the accuracy of the CFO estimate and the accuracy and stability of the underlying clock. Hence, results may vary with the underlying hardware.}

\section{Related Work}\label{sec:related-work}
Despite being an essential metric for localization algorithms and models, there is limited analytical work concerning the variance of the \twr ranging scheme.
For one, Navr{\'a}til and Vejra{\v{z}}ka assess the variance of \twr~\cite{navratil2019bias}.
Based on the original Alt-DS TWR formulation, they assume Gaussian noise on the measured intervals instead of individual reception timestamps. While the employed Alt-DS TWR formula is semantically equivalent to ours, their analysis retains a complex non-linearity and does not discuss more complex NLOS or multipath effects.
They use Taylor Series linearization to approximate the bias and variance of the estimation. Their findings indicate that symmetric response times minimize \twr's variance, which is consistent with our findings.
In addition, their work assumes identical variance for all intervals. 
They assess the adequacy of their model in simulations, but their work does not provide an experimental deployment to verify it under practical conditions.

\n{In contrast, we assume iid noise with individual bias and variance depending on the propagation path; combined with our new formulation, which isolates the clock drift, we can hence analyze the shared nature of timestamps between measured intervals and provide an analytical derivation of the bias and variance even under multi-modal multipath noise, which we verify in an experimental deployment.}

The work by Shalaby et al. derives an analytical model for a different TWR protocol in which the initiator sends two messages initially, and the responder issues the final message~\cite{shalaby2022reducing}.
Hence, their protocol estimates the relative clock using two consecutive messages from the initiator. A longer delay between the two transmissions by the initiator reduces the impact of reception noise on estimating relative clock drift at the price of longer protocol execution. Consequently, their work optimizes response delays regarding the measurement frequency.
An experimental evaluation confirms their theoretical analysis. However, their different TWR protocol results in a substantially different analytical model without overhearing or multipath effects.

For overhearing \td, to our knowledge, our work is the first to derive its variance analytically. Chiasson et al.~\cite{10044978} anticipate a two-fold increase in the variance compared to the TWR exchange but do not include a respective analysis.
As the extraction of \td requires two relative clock mitigations (one for each active party), its variance analysis is slightly more intricate.
Our findings imply that the variance depends on the individual signal path noise. For one, equal variance for all transmission paths results in a five rather than a two-fold increase in variance.

\section{Conclusion}
\label{sec:conclusion}
This work devises a joint analytical model for the bias and variance of \twr and \td under NLOS and multipath effects (see Eq.~\eqref{tof_corrected} and \eqref{tdoa_corrected}). 
As we track the interdependency of timestamping noise and the mitigation of relative clock drift, 
our model relates noise in the underlying timestamping to the bias and variance of \twr and \td.
Our resulting model provides centimeter-accurate predictions and is verified in simulations and experimental UWB deployments with $R^2$ scores of up to 98\%.
Compared to CFO-based single-sided methods, the double-sided mitigation requires one additional message but maintains low error and variance for longer execution times.
\clearpage %

\bibliographystyle{ACM-Reference-Format}
\bibliography{references}

\appendix
\section{Derivations}
\subsection{Deriving the Estimated ToF}\label{appendix:tof_derivation}
\begin{align}
\hattofab^B =& 0.5 \frac{\hat{R}_B^B+\hat{D}_B^B}{\hat{R}_A^A+\hat{D}_A^A} \hat{R}_A^A - 0.5 \hat{D}_B^B \nonumber \\ 
=& 0.5 \frac{R_B^B+\varepsilon_{final}^B+D_B^B-\varepsilon_{poll}^B}{R_A^A+\varepsilon_{resp}^A+D_A^A-\varepsilon_{resp}^A} \hat{R}_A^A - 0.5 \hat{D}_B^B \nonumber \\ 
=& 0.5\frac{k_B}{k_A}\hat{R}_A^A +0.5\frac{\varepsilon_{final}^B-\varepsilon_{poll}^B}{R_A^A+D_A^A}\hat{R}_A^A - 0.5 \hat{D}_B^B \nonumber \\ 
=& 0.5\hat{R}_A^B- 0.5 \hat{D}_B^B + 0.5(\varepsilon_{final}^B-\varepsilon_{poll}^B)\frac{\hat{R}_A^A}{R_A^A+D_A^A} \nonumber \\
=& 0.5 R_A^B - 0.5 D_B^B \nonumber + 0.5\varepsilon_{resp}^B +0.5\varepsilon_{poll}^B \\ &+ 0.5(\varepsilon_{final}^B-\varepsilon_{poll}^B)\frac{\hat{R}_A^A}{R_A^A+D_A^A} \nonumber \\
=& 0.5 R_A^B - 0.5 D_B^B \nonumber + 0.5\varepsilon_{resp}^B +0.5\varepsilon_{poll}^B \\ &+ 0.5(\varepsilon_{final}^B-\varepsilon_{poll}^B)\frac{R_A^A+\varepsilon_{resp}^A}{R_A^A+D_A^A} \nonumber \\ 
\approx& 0.5 R_A^B - 0.5 D_B^B \nonumber + 0.5\varepsilon_{resp}^B +0.5\varepsilon_{poll}^B \\ &+ 0.5(\varepsilon_{final}^B-\varepsilon_{poll}^B)\frac{R_A^A}{R_A^A+D_A^A} \nonumber
\end{align}
\subsection{Deriving TDoA's Second Component}
\label{appendix:tdoa_first_component}
\begin{align}
& \frac{\hat{M}_L^L+\hat{M}_L'^L}{\hat{R}_B^B+\hat{D}_B^B} \hat{D}_B^B \nonumber \\
=& \frac{M_L^B + M_L'^B}{\hat{R}_B^B+\hat{D}_B^B} \frac{k_L}{k_B}\hat{D}_B^B +  \frac{ \varepsilon_{L_3}^L- \varepsilon_{L_1}^L }{\hat{R}_B^B+\hat{D}_B^B} \hat{D}_B^B \nonumber \\
\approx& (1-\frac{\varepsilon_{final}^B-\varepsilon_{poll}^B}{R_B^B+D_B^B})\hat{D}_B^L +  \frac{ \varepsilon_{L_3}^L- \varepsilon_{L_1}^L }{\hat{R}_B^B+\hat{D}_B^B} \hat{D}_B^B \nonumber \\
=& \hat{D}_B^L - \frac{\varepsilon_{final}^B-\varepsilon_{poll}^B}{R_B^B+D_B^B}\hat{D}_B^L +  \frac{ \varepsilon_{L_3}^L- \varepsilon_{L_1}^L }{\hat{R}_B^B+\hat{D}_B^B} \hat{D}_B^B \nonumber \\
\approx& \hat{D}_B^L - \frac{\varepsilon_{final}^B-\varepsilon_{poll}^B}{R_B^B+D_B^B} D_B^L +  \frac{ \varepsilon_{L_3}^L- \varepsilon_{L_1}^L }{\hat{R}_B^B+\hat{D}_B^B} D_B^B \nonumber \\
=& D_B^L-\varepsilon_{poll}^L  - \frac{\varepsilon_{final}^B-\varepsilon_{poll}^B}{R_B^B+D_B^B} D_B^L +  \frac{ \varepsilon_{L_3}^L- \varepsilon_{L_1}^L }{\hat{R}_B^B+\hat{D}_B^B} D_B^B \nonumber \\
\approx& D_B^L-\varepsilon_{poll}^L  - \frac{\varepsilon_{final}^L-\varepsilon_{poll}^L}{R_B^B+D_B^B} D_B^B +  \frac{ \varepsilon_{L_3}^L- \varepsilon_{L_1}^L }{R_B^B+D_B^B} D_B^B \nonumber
\end{align}

\end{document}